\documentclass{vgtc}                          

\graphicspath{{figures/}{pictures/}{images/}{./}} 

\usepackage{times}                     

\usepackage{tabu}                      
\usepackage{booktabs}                  
\usepackage{lipsum}                    
\usepackage{mwe}                       

\usepackage{mathptmx}                  

\usepackage{bm}

\usepackage{enumitem}
\usepackage{amsmath}

\usepackage[T1]{fontenc}

\newcommand{\revision}[1] 
{\textcolor{black}{#1}}

\newcommand{\cameraRevision}[1] 
{\textcolor{black}{#1}}
\onlineid{0}

\vgtccategory{Research}

\vgtcinsertpkg

\title{
What if Virtual Agents Had Scents? Users' Judgments of Virtual Agent Personality and Appeals in Encounters 
}

\author{Dongyun Han\thanks{e-mail: dongyun.han@usu.edu}\\ %
        \scriptsize Utah State University %
\and Siyeon Bak\thanks{e-mail: 123siyeon84@gmail.com}\\ %
     \scriptsize Hallym University %
 \and So-Hui Kim\thanks{e-mail: kimsohee12360@gmail.com}\\ %
 \scriptsize Hallym University %
 \and Kangsoo Kim\thanks{e-mail: kangsoo.kim@ucalgary.ca}\\ %
 \scriptsize University of Calgary%
  \and Sun-Jeong Kim\thanks{e-mail: sunkim@hallym.ac.kr}\\ %
 \scriptsize Hallym University %
\and Isaac Cho\thanks{e-mail: isaac.cho@usu.edu (correspondence)}\\ %
\scriptsize Utah State University 
     }

\teaser{
  \centering
  \includegraphics[width=.9\linewidth]{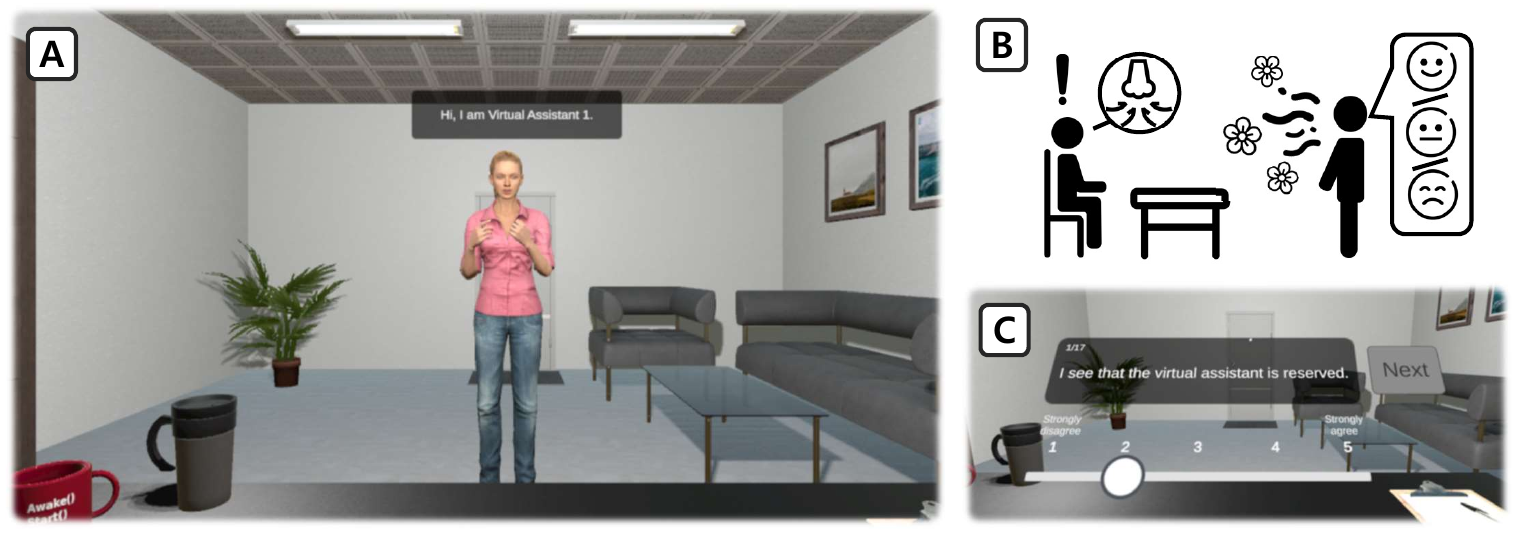}
  \caption{(A) Participants meet virtual agents in a virtual office. (B) This work evaluates the effect of virtual agents' non-verbal cues, including olfactory cues, emotional expressions, and gender, on participants' judgment of virtual agents. (C) Participants evaluate the impressions of virtual agents in terms of Big 5 personality traits and Appeal metrics in VR.}
  \label{fig:teaser}
}

\abstract{
    Incorporating multi-sensory cues into Virtual Reality (VR) can significantly enhance user experiences, mirroring the multi-sensory interactions we encounter in the real-world. 
    Olfaction plays a crucial role in shaping impressions when engaging with others.
    This study examines how non-verbal cues from virtual agents—specifically olfactory cues, emotional expressions, and gender—influence user perceptions during  encounters with virtual agents. 
     Our findings indicate that in unscented, woodsy, and floral scent conditions, participants primarily relied on visually observable cues to form their impressions of virtual agents. 
     Positive emotional expressions, conveyed through facial expressions and gestures, contributed to more favorable impressions, with this effect being stronger for \cameraRevision{the female agent than the male agent. 
     }
     However, in the unpleasant scent condition, participants consistently formed negative impressions, which overpowered the influence of emotional expressions and gender, suggesting that aversive olfactory stimuli can detrimentally impact user perceptions. 
     Our results emphasize the importance of carefully selecting olfactory stimuli when designing immersive and engaging VR interactions. Finally, we present our findings and outline future research directions for effectively integrating olfactory cues into virtual agents. 
} 


\begin{document}

\firstsection{Introduction}

\maketitle


Human senses play a pivotal role in shaping judgments when interacting with others~\cite{ambady2008first, spence2021scent}.
Individuals typically aim to create positive impressions, as these are often difficult to alter once established~\cite{Muthukrishnan2007just}.
This concept extends to virtual environments, where forming positive impressions of virtual agents is essential for fostering and maintaining user engagement in \revision{Virtual Reality (VR) applications~\cite{bailenson2005independent}.}
Virtual agents are commonly defined as computer-controlled characters that appear to be virtual humans. Many VR applications employ virtual agents as instructors or collaborators to provide narrative experiences for users~\cite{ranjbartabar2019first, cafaro2012first}. 
Earlier research demonstrated that virtual agents' visual cues, such as emotional facial expressions, gestures, and gender, are strongly related to the quality of user satisfaction and social interactions with virtual agents~\cite{shang2020effects}. 
According to psychological research~\cite{ambady2008first, spence2021scent, argyle2013bodily}, olfactory cues also play a crucial role in judging others. However, our understanding of virtual agent perception is limited to these visual cues, often overlooking the potential contributions of other sensory modalities, particularly olfaction, even though psychological research has shown that olfactory cues also play a crucial role in social evaluation and interpersonal judgments.


Researchers have explored the integration of multi-sensory modalities into VR to enhance the user experience~\cite{ranasinghe2018season, ebrahimi2016empirical, hagelsteen2017faster}. 
Studies have demonstrated that multi-sensory integration can enhance user experiences, leading to a greater sense of presence and improved perception of virtual environments and objects.
While much of such work has focused on auditory and haptic feedback, olfactory cues can also offer a distinct opportunity to enrich VR experiences by eliciting emotions~\cite{chanes2016redefining, sullivan2015olfactory} and improving cognitive functions~\cite{amores2017essence, holland2005smells}. 
For instance, similar to how the smell of coffee activates olfactory representations in the brain, olfactory cues in VR could increase the salience of virtual objects, aiding in their recognition and enhancing the sense of presence~\cite{degel2001implicit, persky2020olfactory}. 
Research on incorporating olfaction in VR remains relatively unexplored, yet it has the potential to meaningfully influence users’ judgments of virtual agents, either positively or negatively.

This paper investigates the effect of olfactory cues along with visual cues, specifically emotional expressions, and gender, on users' impressions of virtual agents.
Earlier research has found that users generally prefer positive-looking and female virtual agents, leading to higher acceptance and reduced discomfort during interactions~\cite{volonte2020effects, sajjadi2020effect}. 
However, the effect of olfactory cues on the perception of virtual agents has not yet been examined. 
Understanding the interaction effects of these characteristics is also crucial~\cite{arboleda2024beyond}. 
This research gap could be attributed to technical challenges, such as quantitatively controlling the extent of scent exposure, despite the introduction of various olfactory devices at affordable prices in previous works \cite{javerliat2022nebula, myung2023enhancing}. 
Recognizing these technological constraints, we design and perform a user study to examine the influence of the three non-verbal cues from virtual agents on users'  impressions, including olfactory cues, emotional expressions, and gender of virtual agents.
We \revision{assess and report} their effects using both quantitative biosignal measurements and participants' subjective evaluations based on the Big Five personality traits and appeal metrics. 
Finally, we discuss the implications of \revision{our} findings, acknowledge the limitations of our study, and outline future research directions for the effective integration of olfactory cues into virtual agents.


\section{Related Work}
\subsection{Benefits of Olfactory Cues in VR}

Earlier researchers have investigated how olfaction can be integrated into VR~\cite{niedenthal2023graspable}. 
They have introduced a variety of olfactory devices, typically by diffusing vapors~\cite{ranasinghe2018season, javerliat2022nebula, myung2023enhancing} and heating solids~\cite{dobbelstein2017inscent} or liquids~\cite{covington2018development}.
A design space~\cite{maggioni2020smell} for olfactory experience includes chemical, emotional, spatial, and temporal features. These features refer to scent types, receiver emotional responses, spatial sources of scents, and scent duration, respectively.
In the real-world, these features enable us to actively explore space and make judgments about the identity, concentration, and combination of scents~\cite{amores2017essence, dobbelstein2017inscent}.
However, replicating olfactory experiences in VR is challenging due to technical limitations, such as controlling scent intensity and providing a wide variety of scents. Nonetheless, understanding its benefits is not trivial but essential for providing better VR user experiences~\cite{covarrubias2015vr, tsai2021does}.


Previous research has shown that olfactory cues can enhance sense of immersion and the perceived presence of virtual objects in VR~\cite{munyan2016olfactory, persky2020olfactory}. 
For instance, Persky and Dolwick~\cite{persky2020olfactory} demonstrated that the scent of french fries heightened user immersion in a virtual restaurant. 
\revision{Tsai et al.~\cite{tsai2021does}} investigated the correlation between visual and scent perceptions based on their congruency and intensity. 
They suggested that semantically congruency between vision and olfaction allows users to process them more quickly and accurately. 
Baus and Bouchard~\cite{baus2017exposure} investigated how pleasant (apple/cinnamon) and unpleasant (urine) scents affected user presence in VR. 
Interestingly, they discovered that participants who were exposed to the unpleasant scent had a stronger sense of presence than those who were exposed to the pleasant scents.

To the best of our knowledge, only one study has investigated the effect of olfactory cues on virtual agents. Quintana et al.~\cite{quintana2019effect} examined the impact of three scents—fear-related, joy-related, and neutral—on participants' interactions in a virtual environment. The fear-related and joy-related scents were derived from sweat samples collected from individuals watching horror films and funny animal videos, respectively, while the neutral scent consisted of sterile gauze. 
Their findings revealed that participants exposed to fear-related body odorants experienced significantly higher anxiety and exhibited lower trust toward virtual agents compared to those exposed to joy-related or neutral scents. 
However, no significant differences were observed in participants' reactions to joy-related and neutral scents.
While olfactory cues can enhance virtual experiences, their specific influence on perceptions of virtual agents, particularly regarding personality traits and appeal, remains unclear.
\subsection{Emotional Expressions and Gender of Virtual Agents}\label{sec_rw_virtualAgents}



Earlier works demonstrated that virtual agents' verbal~\cite{ter2010turn, sallnas2005effects} or non-verbal~\cite{gunawardena1997social, takashima2008effects} cues are strongly related to user satisfaction and the quality of social interactions between users and  agents.
While various cues have been studied, including voice~\cite{higgins2022sympathy}, rendering style~\cite{mcdonnell2012render}, and head/gaze motion~\cite{takashima2008effects}, this section focuses on how agents' emotional expressions and gender influence users' experiences when interacting with them in VR.

Virtual agents can convey emotional expressions through facial expressions and gestures. 
Earlier research demonstrated that users respond more positively and experience higher social presence when agents display positive emotional expressions compared to negative ones. 
For example, Bönsch et al.~\cite{bonsch2020impact} examined interpersonal distance between participants and virtual agents expressing happy, neutral, or angry emotions to assess perceived intimacy with the agents. They found that anger agents elicited greater distance, indicating less intimacy. 
Cafaro et al.~\cite{cafaro2012first} also explored the effect of virtual agents' non-verbal cues in a museum guide scenario, showing that smiles and eye contact foster positive first impressions and socio-emotional connections. 
Volonte et al.~\cite{volonte2020effects} conducted a study using a virtual agent crowd model with rich features, including eye gaze, facial expressions, and body motion. They found that positive emotional crowds resulted in the highest interaction counts and the longest interaction times between participants and virtual agents.
These studies suggest that incorporating positive expressions in virtual agents enhances user engagement and leads to stronger connections during interactions~ 
which is known as emotional contagion between users and virtual agents, leading to stronger connections during interactions~\cite{tsai2012study}.



The gender of virtual agents has also been studied. Nag and Yal{\c{c}}{\i}n~\cite{nag2020gender} found that virtual agents' visual appearances do not introduce gender bias in participants' recognition of their gender. However, previous studies have shown that participants' perceptions of a virtual agent's gender significantly impact their VR experience.
Sajjadi et al.~\cite{sajjadi2020effect} investigated the impact of a virtual instructor's gender in an educational virtual field trip. 
They found that the female virtual instructor elicited higher levels of presence and was perceived by participants as more effective in facilitating learning. 
Regal et al.~\cite{regal2022marcus} investigated the gender effect in a training scenario and reported that their participants rated a female agent as more organized and productive than a male agent.
Shang et al.~\cite{shang2020effects} had participants perform a customized task with male or female virtual agents. While there was no significant difference in task error rates between genders, participants found the female agent more appealing.
Similar preference was observed in proxemics research involving virtual agents in VR.
Bailenson et al.~\cite{bailenson2003interpersonal} investigated interpersonal distances between users and virtual agents to determine their level of intimacy. They reported that participants felt higher intimacy to a female agent having smaller interpersonal distance than a male agent. An emotional bias that people associate more positive attributes with females when compared to males is known as \textbf{``women-are-wonderful effect''} in psychological and sociological research~\cite{eagly1994people, eagly1991women}. 
Building on these findings, we further examine how olfactory cues affect users'  impressions of virtual agents and explore their interaction with emotional expressions and gender.

\section{User Study}
\subsection{Research Motivation and Questions}

\begin{figure*}[t]
  \centering 
  \includegraphics[width=0.8\textwidth]{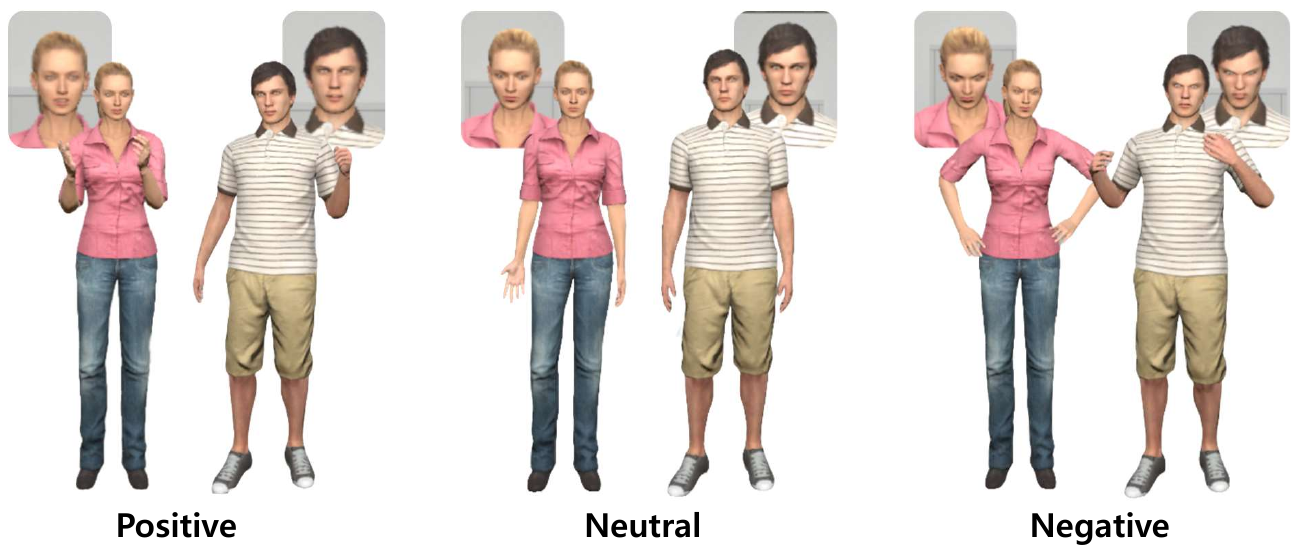}
  \caption{Within-subject factors: emotional expression and gender. This study considers three emotional expressions, including \textit{Positive}, \textit{Neutral}, and \textit{Negative,} and two genders of \textit{Female} and \textit{Male} virtual agents. 
  }
  \label{fig:gender three expression}
  \vspace{-.5cm}
\end{figure*}

Designing appealing virtual agents that enhance intimacy and narrative experiences is central to increasing user engagement in VR applications~\cite{ranjbartabar2019first, cafaro2012first}. 
Given that impressions formed during initial encounters are often difficult to change, it is important to understand the factors that shape users' judgments of virtual agents' characteristics and the implications of those judgments ~\cite{sunnafrank2004first, riggio1986impression, cafaro2012first}.
Previous real-world studies~\cite{sorokowska2012does, sorokowska2016body} have shown that body odor conveys personality traits, underscoring the significant role of olfactory cues in impression formation. 
With the increasing use of virtual agents in immersive environments, understanding the impact of olfactory cues on users' judgments of the virtual agents is essential for creating more realistic and socially engaging virtual interactions. 
Focusing on olfaction as the primary modality of interest, this study investigates how olfactory cues interact with other non-verbal cues, namely emotional expressions, and gender, to shape users' judgments of virtual agents. We aim to examine the relative and combined effects of these cues on perceiving personality trains and social appeals.

Our research questions (RQs) are as follows:
\vspace{-.2cm}
\begin{enumerate}
[label=\textbf{RQ\arabic*.}, leftmargin=0.35in]
\setlength\itemsep{0pt}

\item To what extent do \emph{olfactory cues} affect users' judgments of virtual agents’ characteristics and attractiveness?

\item How do \emph{olfactory cues} interact with non-verbal cues, specifically \emph{gender, emotional expressions,} to influence users' judgments of virtual agents? 


\end{enumerate}

\subsection{Study Design and Independent Variables}

We employ a mixed-design experiment to investigate three non-verbal cues of virtual agents: olfactory cues (body scent), emotional expressions (facial and gestural), and gender. 
The olfactory cue is treated as a between-subject factor, while emotional expressions and gender are within-subject factors. 
We selected olfactory cues as the between-subject factor for two reasons: 1) to prevent olfactory fatigue and ensure participants' senses remained consistent throughout the study, and 2) because it is difficult to maintain equal olfactory intensity across different scents. Therefore, we prioritize reliable scent delivery over intensity matching, making the scent condition a between-subject factor. 

\paragraph{Olfactory Cue:} We considered four conditions including \textit{No Scent, Woodsy, Floral,} and \textit{Unpleasant}.
These scented conditions (pure air, pleasant scents, and unpleasant scents) are often explored together in previous psychology studies~\cite{dematte2007olfactory, cook2018simultaneous, bensafi2002modulation} investigating the effects of olfactory cues on human judgment in real-world settings.
Each scent is presented to participants when they encounter a virtual agent in the VR environment.


\textit{No Scent} is a baseline condition in which the virtual agent emits no scent. 
\textit{Woodsy} and \textit{Floral} scents are represented by commercially available teakwood and Japanese cherry blossom fragrances, respectively \footnote{Teakwood and Japanese Cherry blossom by Bath \& Body Works. In March 2025, they are ranked 12th and 46th, respectively, among the best-selling men's and women's body sprays (\url{https://www.amazon.com/Best-Sellers-Men's-Scented-Body-Sprays/zgbs/beauty/16262034011} and \url{https://www.amazon.com/Best-Sellers-Women's-Body-Sprays-Fragrance/zgbs/beauty/3783161)}.}.
These are selected based on their popularity on Amazon in 2023 as the top-selling perfume and fragrance for men and women. 
Next, \textit{Unpleasant} is operationalized using a bad gas stench, intended to simulate a negative olfactory experience \footnote{Laughing Smith's - Wet Farts}. The products used in this work are shown in Fig.~\ref{fig:apparatus}A.



\paragraph{Emotional Expression:} We examined three levels of emotional expression: \textit{Neutral}, \textit{Positive}, and \textit{Negative}.
We use Microsoft Rocketbox virtual human animations (neutral, excited, and angry walking and talking gestures)~\cite{gonzalez2020rocketbox} as shown in Fig~\ref{fig:gender three expression}. 
In the \textit{Neutral} condition,  virtual agents have a face with no specific emotional cues and have minimal body gestures.
In the \textit{Positive} condition, virtual agents have a smiling face and exhibit gentle walking and talking gestures, whereas in the \textit{Negative} condition, agents display an angry face and aggressive gestures.
The virtual agent's head is always oriented toward the participants to ensure its face remains visible at all times.

\paragraph{Gender:} We considered two genders of virtual agents (\textit{Female} and \textit{Male}) implemented using male and female characters from the Microsoft Rocketbox library (Female\_Adult\_01 and
Male\_Adult\_01). Fig.~\ref{fig:gender three expression} depicts two virtual agents displaying three different emotional expressions.

To eliminate uncontrolled factors, such as appearance, which can influence participants' impressions~\cite{bailenson2001equilibrium, nowak2003effect}, we used virtual agents with the same face and outfit across all task trials. This approach was also used in previous research to isolate the effects of other factors on users' judgment of virtual agents~\cite{cafaro2012first}.


\subsection{Task}

In each task, a participant encounters one virtual agent in a virtual office (Fig.~\ref{fig:teaser}). 
At the beginning of each task, the office door opens and a virtual agent enters. The agent stops approximately 2.5 meters from the participant maintaining a social distance~\cite{hall1963system}. When the agent stops, the olfactory device is activated and begins to emit a scent. 
The agent displays a text box introducing itself for 7 seconds on top of its head, with the message ``I am virtual agent \textit{\#X}.'' \textit{\#X} represents the task number. The agent has no voice. 
While it is known that impression of others can be formed after 100 ms exposure to a human face~\cite{willis2006first}, we decide this 7 seconds to expose scents in the following two reasons. 
First, we chose a 7-second exposure to ensure that participants had enough time to recognize the scent, which typically requires at least 1.0 second~\cite{olofsson2014time}, while also avoiding nose blindness, which can occur after prolonged exposure (e.g., around 15 seconds~\cite{stortkuhl1999olfactory}). Nose blindness is is a temporary condition where a person loses their ability to detect certain smells, also known as the olfactory fatigue and adaptation. 
Since it can occur due to various factors, such as scent intensity, we set the exposure duration within this safe range.
Second, 7 seconds is determined based on the 7/11 rule, a belief that people could make 11 judgments about others (e.g., gender, perceived credibility, trustworthiness, etc) in 7 seconds. However, this rule lacks systematic empirical validation. 
After this brief introduction, the agent turns around, opens the door, and leaves. Following that, the participant completes a questionnaire within the virtual office. The questionnaire is discussed in the following section. It includes a total of 16 statements to evaluate. Each statement is popped up in front of the participant as shown in Fig.~\ref{fig:teaser}C. The participant can evaluate the statement by pointing a score in the slider with a ray emitted by one VR controller and pressing its trigger button. The participant can proceed to the next statement by clicking the `Next' button.
We asked the participant to pay attention to and recall the agent's non-verbal cues while meeting with the virtual agent and answering the questionnaire.
After completing the questionnaire, participants take a 5-minute break with ventilated air to clear any lingering scents. This break helps them recover from potential nose blindness. 

\subsection{Measurements and Hypothesis}
For each task trial, participants complete a questionnaire in VR after encountering a virtual agent. 
The questionnaire has 16 statements assessing participants' perceived attractiveness (5), personality traits (10), and willingness to engage in future interactions (1). For a more detailed list of questions, see the supplementary material.
We also record their biosignal before and during the task to objectively analyze the effect of the virtual agent's non-verbal cues.


\paragraph{Appeal:} 
It is assessed on a 7-point Likert scale across five metrics: affinity, credibility, emotional comfort, realism, and social presence (sense of being together). 
They are chosen based on previous research~\cite{cafaro2012first, bailenson2005digital, zibrek2019photorealism}. 
They are comparable to five conceptual indices used in the Godspeed Questionnaire~\cite{bartneck2009measurement}, widely utilized in human-robot/agent interaction research.

\paragraph{Big Five Personality Trait:}
The Big Five personality traits model was employed to evaluate virtual agents~\cite{neff2011don, krenn2014effects, doce2010creating}.
The five traits are agreeableness, extraversion, conscientiousness, neuroticism, and openness. Agreeableness is a behavioral trait that describes how one treats others. 
Extraversion reflects how much people enjoy interacting with others. 
Conscientiousness denotes the quality of being cautious or diligent. Neuroticism is a trait that determines how frequently a person experiences negative emotions in daily life. 
Finally, openness is how individuals evaluate and respond to new experiences. 
We use a brief 10-item Big Five questionnaire~\cite{rammstedt2007measuring}, where participants rate each statement on a 5-point Likert scale (1: strongly disagree to 5: strongly agree). Statements are weighted positively or negatively (+1 or -1) to assess traits.

\paragraph{Future Engagement:} This binary question asks participants to respond with `Yes' or `No' to indicate whether they would be willing to interact with virtual agents in the future.

\paragraph{Biosignal:} 
PPG (photoplethysmography) is a widely used indicator of physiological and emotional arousal in user responsiveness studies~\cite{lim2023effects, lyzwinski2023use, shaffer2017overview}. 
PPG records blood volume pulse (BVP), measuring heart rate variability (HRV) by using optical plethysmography. pNN50 (\%) is an HRV measurement, the variation in time between successive heartbeat intervals greater than 50 ms.
It may differ between individuals. Therefore, we analyze the changes ($\Delta$pNN50 ppt (percentage point)) between pNN50 values after exposure to each virtual agent condition and the baseline value measured before the first task. 
A positive change can be interpreted as participants forming a favorable impression of  virtual agents, while a negative change suggests an unfavorable impression.

\begin{figure}[t]
  \centering 
  \includegraphics[width=.95\linewidth]{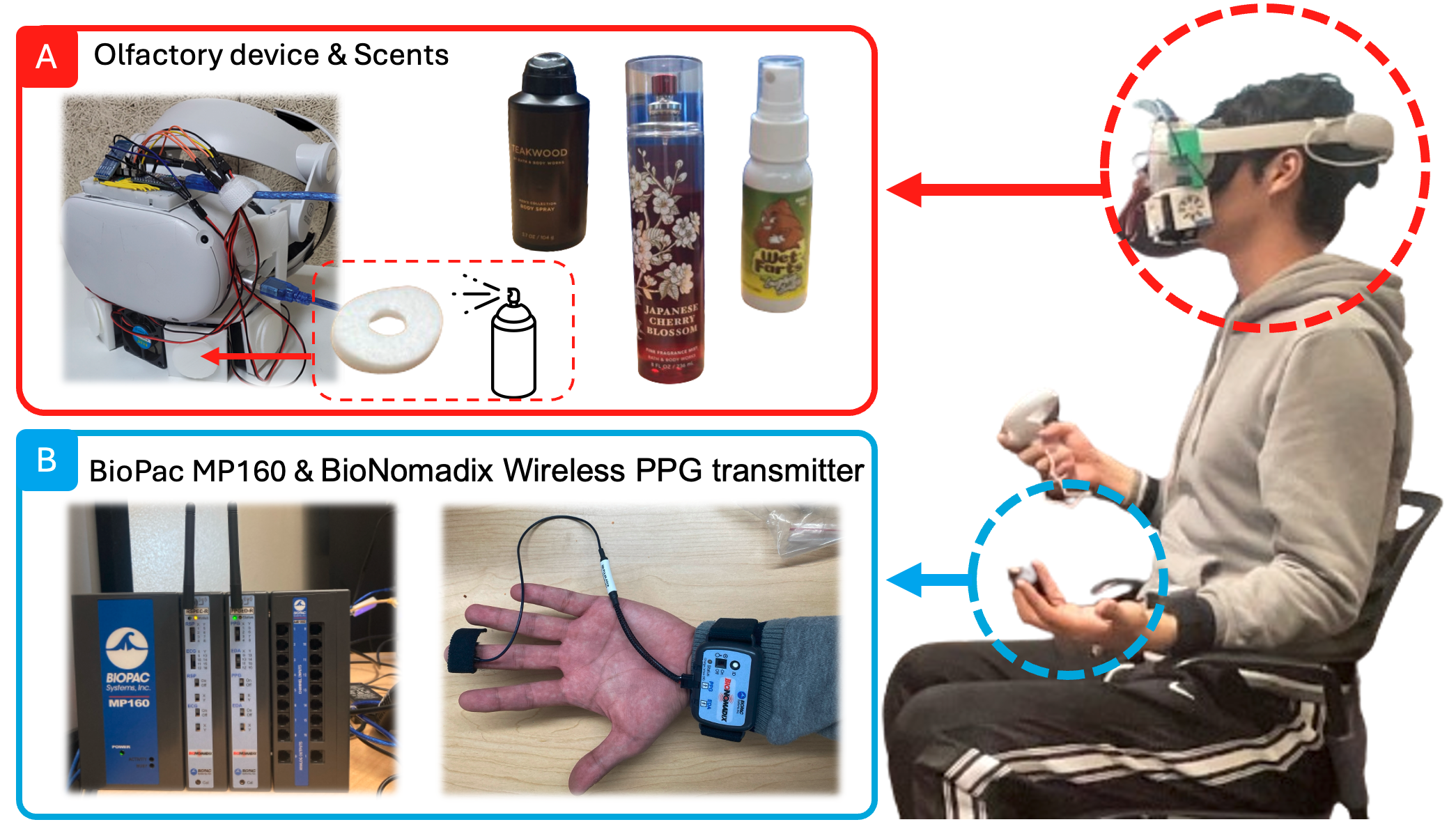}
  \caption{
  (A) An olfactory device with three distinct scents is used in this study. The three different scents (\textit{Woodsy}, \textit{Floral}, and \textit{Unpleasant} from left to right) plus the \textit{No Scent} condition are the between-subject factors. (B) The BioPac and its wireless PPG transmitter are used to monitor and evaluate participants' biosignal changes when they encounter virtual agents in VR.}
  \label{fig:apparatus}
  \vspace{-.5cm}
\end{figure}


\begin{enumerate}
[label=\textbf{H\arabic*.}, leftmargin=0.25in]
\setlength\itemsep{0pt}
\item The \textit{Woodsy} and \textit{Floral} scents will give participants a more \revision{positive impression (i.e., lower neuroticism and higher levels of the remaining traits and appeal metrics)} of virtual agents than \textit{No Scent} and \textit{Unpleasant}. This is because they provide additional positive non-verbal cues to participants.

\item \revision{The \textit{Unpleasant} scent} is expected to negatively affect impressions of virtual agents \revision{(higher neuroticism and lower levels of the remaining traits and appeal metrics)}, as it may cause discomfort in encounters.

\item The scented conditions \revision{(\textit{Woodsy}, \textit{Floral}, and \textit{Unpleasant})} are expected to enhance the social presence and realism of virtual agents, as prior research indicates that olfactory cues \revision{increased} the perceived presence of virtual objects in VR~\cite{persky2020olfactory, baus2017exposure}.

\item Virtual agents accompanied by scents that align with real-world gender-based scent associations (i.e., woodsy scents with a male agent, floral scents with a female agent) are expected to be more appealing than mismatched combinations, as such congruence may enhance users’ sense of familiarity and \revision{semantic coherence~\cite{tsai2021does}}.


\end{enumerate}

\noindent The following hypotheses are grounded in earlier research findings, as reported in Sec~\ref{sec_rw_virtualAgents}~\cite{bonsch2020impact, brady1978interpersonal, polit1977sex}.

\begin{enumerate}[label=\textbf{H\arabic*.}]
\setcounter{enumi}{4}
\setlength\itemsep{0pt}
\item Virtual agents with \revision{the positive emotional expression} will be perceived more favorably than those with \textit{Neutral} or \textit{Negative}, as they are expected to enhance participant relaxation and evoke positive feelings.


\item \textit{Female} agents will be perceived more positively than \textit{Male} agents based on the women-are-wonderful effect. 

\end{enumerate}

\subsection{Apparatus}\label{sec_apparatus}

This study employs three main apparatuses: a VR headset and one of its controllers, an olfactory device, and a biosignal recording device. We used Meta Quest 2, which provide a horizontal Field of View (FoV) of 104$^\circ$ and a vertical FoV of 98$^\circ$, with a resolution of 1832 × 1920 pixels per eye. 
The virtual scene (Fig.~\ref{fig:teaser}A) was created in Unity 2021.3.19f1 and ran on a Windows 11 desktop with an Intel Xeon W-2245 CPU (3.90GHz), 64GB RAM, and an Nvidia GeForce RTX 3090 graphics card.

Fig.~\ref{fig:apparatus}-A shows the olfactory prototype device\footnote{3D print files for the prototype device are available in \url{https://anonymous.4open.science/r/Wearable_Olfactory_Device-9A86}}. It attaches to the Meta Quest 2 and operates at 5V, charging via a connection to the HMD's USB-C port. It is low-cost (approximately \$50) and lightweight (250g)~\cite{myung2023enhancing}.  
It has five fans, one of which is positioned in front of the user's nose for ventilation, and the remaining fans can contain cotton soaked in scented liquid, providing the user with up to four scents. In this study, the ventilation fan and one scent fan located to the right of the ventilation fan are used. Only one of the fans is activated when the device is turned on, running at full speed throughout the entire duration while the virtual agent is present or during the ventilation stage.
To clearly deliver the scents, \textit{Woodsy} and \textit{Floral} scents are sprayed onto cotton pads four times, \revision{whereas the \textit{Unpleasant} odor is applied only once due to its notably stronger intensity.}
\revision{While the exact scent strength is not precisely measurable, this adjustment is intended to approximately balance their perceived intensity.}
Dispersing scents is controlled by an Arduino Nano connected to a computer running Unity via Bluetooth (HC06 Bluetooth Serial Module). For more information, please refer to \cite{myung2023enhancing}.



To record biosignal, the BioPac MP160\footnote{\url{https://www.biopac.com/product-category/research/systems/mp150-starter-systems/}} amplifier system and the BioNomadix wireless PPG transmitter device are used (Fig.~\ref{fig:apparatus}-B).
The PPG sensor, called BioNomadix Pulse Transducer, is attached to the participant's middle finger, and the wireless transmitter is worn on the participant's non-dominant wrist.

\subsection{Participants}
We recruited 23 participants (14 female and 9 male, average age of 20.1, ranging from 18 to 24) through SONA at our university. All participants had (corrected) 20/20 vision.
Participants were compensated with 2.0 SONA credits as a reward (IRB \#13909). 
A priori power analysis was conducted using G*Power with the following parameters: effect size=0.4, power=0.95, number of groups=4, number of measures=6, and other default settings. The results indicated that a minimum sample size of 20 participants was required for the study.
Participants were assigned to the between-subject conditions as follows: \textit{No Scent} (6 participants; 4 females and 2 males) \textit{Woodsy} (6 participants; 4 females and 2 males), \textit{Floral} (6 participants; 3 females and 3 males), and \textit{Unpleasant} (5 participants; 3 females and 2 males).


\subsection{Procedures}
The study took approximately 60 minutes. All procedures are performed while participants are seated on a real-world chair, which is synchronized with a chair in the virtual office.
Upon arrival, participants sign an informed consent form and complete a demographic questionnaire. 
An instructor then provides an overview of the study, including its objectives, procedures, virtual agents, and the devices used. 
Participants are informed that they will encounter a total of six distinct virtual agents (2 genders \texttimes{} 3 emotional expressions) one at a time. \revision{They are also instructed to treat an olfactory cue as if it is coming from the virtual agent during each encounter.}
For each agent, participants will evaluate personality traits, appeal, and their willingness for future engagement. 
They also learn how to operate the VR controller to answer survey questions in VR.
After this briefing, the instructor attaches the biosignal device to participants, as illustrated in Fig.~\ref{fig:apparatus}-B.
Finally, participants put on the VR device, which is connected to the olfactory device, and prepare to begin the main session.

During the main session, participants complete the task six times.
The virtual agents were presented in a randomized order. 
Before starting the first task, participants take a 5-minute break. During this period, the instructor records their biosignal, which serves as a baseline for comparison with biosignals collected in subsequent tasks.
After the break, the instructor attaches a scent source to the olfactory device, and participants begin the first task. 
The biosignal is measured for approximately 2 minutes as participants encounter the virtual agent and complete the questionnaire.
Following the task, the instructor removes the scent source from the olfactory device and instructs participants to take another 5-minute break while the device is ventilated.
This break helps participants recover from any biosignal fluctuations and nose blindness.
After the break, participants proceed to the next task, repeating the process.

After completing all tasks, participants fill out a post-questionnaire to report subjective feedback on how the virtual agents' nonverbal cues influenced their judgment. 
At the end of the study, the instructor ventilates the room and sanitizes the olfactory device with alcohol to eliminate any residual scents.

\subsection{\revision{Analysis Methods}}

\revision{
For the analysis, we conducted a mixed-design ANOVA in SPSS (version 29.0.1.0) with a 95\% confidence level. 
While ANOVA is typically used for continuous variables, prior research ~\cite{norman2010likert, johnson1983ordinal} suggests that it can be applied to Likert-type data with five or more categories. 
Our Likert items satisfy this requirement.
We applied the Greenhouse-Geisser correction to account for potential violations of the sphericity assumption.
Pairwise comparisons were conducted following significant ANOVA effects.
}

\bgroup
\def\arraystretch{1.2}
\begin{table}[t]
\centering
\small
\caption{\label{table:appeal_results} \revision{Appeal Results: Significant interaction and main effects}}
\begin{tabular} {m{1.2cm} | m{2cm} | m{1.65cm} | m{0.7cm}| m{0.7cm}}
\toprule
\textbf{} & \textbf{Factor}  &  {$\bm{F}$} & {$\bm{p}$} & {$\bm{\eta_p}^2$}\\
\midrule
   & {Olfactory Cue \newline \texttimes{} Emot. Expression}     &  {F(3, 19)=7.819} & {.025}  &  {.552}  \\ \cline{2-5}
  \textbf{Affinity}   & {Emot. Expression \newline \texttimes{} Gender}   &   {F(2, 38)=4.379}  & {.009}  & {.388}    \\ \cline{2-5}
    &  {Olfactory Cue}      &  {F(3, 19)=9.628}  & {<.001}  & {.603}   \\ \cline{2-5}
   &  {Emot. Expression}      &  {F(2, 38)=53.011}  & {<.001}  & {.736}   \\
\midrule
                    & {Olfactory Cue}      &  {F(3, 19)=5.013}  &  {.010} &  {.442}  \\ \cline{2-5}
\textbf{Credible}   & {Emot. Expression}      &  {F(2, 38)=22.728} &  {<.001} &   {.545} \\ \cline{2-5}
                    & {Gender}      & {F(1, 19)=6.566}  & {.019}  &  {.257} \\ 
\midrule
\textbf{Emotional \newline Comfort}   & {Emot. Expression}      &  {F(2, 38)=16.843}  &  {<.001} &  {.470}  \\ 
\midrule
\textbf{ }   & {Emot. Expression \newline \texttimes{} Gender}      &  {F(2, 38)=4.554}  &  {.017} &  {.193}  \\ \cline{2-5}
\textbf{Realism}  & {Olfactory Cue}      &  {F(3, 19)=3.958} &  {<.001} &   {.385} \\ \cline{2-5}
                & {Emot. Expression}      &  {F(2, 38)=3.776}  & {.032}  &  {.166}  \\ 
                \midrule
\textbf{Social \newline Presence}   & {Gender}      &  {F(1, 19)=4.621}  &  {.045} &  {.196}  \\ 

\bottomrule
\end{tabular}
\vspace{-2ex}
\end{table}

\section{Results}



\subsection{Appeal Analysis Results} \label{Sec:AppealAnalysisResults}


\begin{figure}[t]
  \centering 
  \includegraphics[width=.8\linewidth]{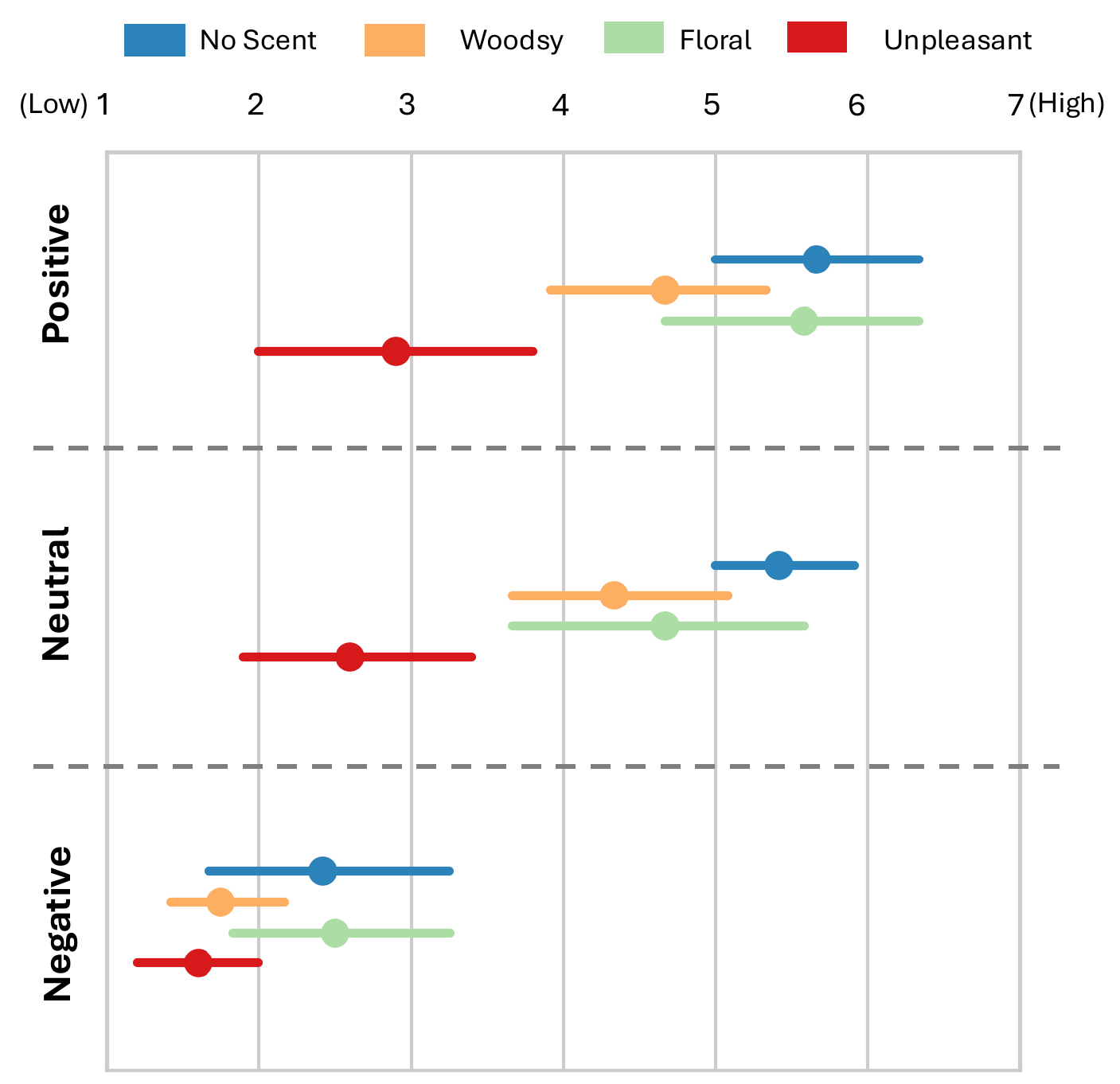}
  \vspace{-.3cm}
  \caption{Appeal - Affinity results ranging from 1 to 7, with the error bars representing the 95\% confidence interval. We found an interaction effect between emotional expression and olfactory cue.
  }
  \label{fig:appeal-affinity}
  \vspace{-.5cm}
\end{figure}

\revision{Table~\ref{table:appeal_results} presents the observed interaction and main effects in the  Appeal results.}

\paragraph{Affinity:} Two interaction effects were reported between emotional expressions and olfactory cue ($p$=.025).
Simple effects on the olfactory cue were found when interacting with \textit{Neutral} ($p$\textless.001, $F(3, 45)$=8.869, $\eta^2_p$=.388) and \textit{Positive} ($p$\textless.001, $F(3, 45)$=8.092, $\eta^2_p$=.366) agents as shown in Fig~\ref{fig:appeal-affinity}. 
In \textit{Neutral}, \textit{Unpleasant} ($m$=2.600 [1.695, 3.504]) had a clearly lower affinity score than \textit{No Scent} ($m$=5.417 [4.917, 5.920], $p$\textless.001), \textit{Woodsy} ($m$=4.333 [3.462, 5.204], $p$=.030), and \textit{Floral} ($m$=4.667 [3.507, 5.827], $p$=.006). 
Similarly, when interacting with \textit{Positive} virtual agent, \textit{Unpleasant} ($m$=2.900 [1.810, 3.990]) has a lower affinity score than \textit{No Scent} ($m$=5.667 [4.885, 6.449], $p$\textless.001), \textit{Woodsy} ($m$=4.667 [3.839, 5.494], $p$=.033), and \textit{Floral} ($m$=5.583 [4.589, 6.578], $p$\textless.001).  In both conditions, we found no clear difference between \textit{No Scent}, \textit{Woodsy}, and \textit{Floral}. 

In addition, there was an interaction effect between emotional expression and gender ($p$=.009). We found a simple effect on gender with the \textit{Positive} virtual agent ($p$=.003, $F(1, 22)$=11.055, $\eta^2_p$=.334), showing \textit{Female} ($m$=5.217 [4.436, 5.999]) had a higher score than \textit{Male} ($m$=4.348 [3.663, 5.033]).

We found a main effect on olfactory cue ($p$\textless.001). 
The average affinity values of olfactory cues were as follows: \textit{No Scent}=4.5 [3.892, 5.108], \textit{Woodsy}=3.583 [2.976, 4.191], \textit{Floral}=4.250 [3.642, 4.858], and \textit{Unpleasant}=2.367 [1.701, 3.032]. Pairwise comparisons disclosed that \textit{Unpleasant} was statistically lower than \textit{No Scent} ($p<0.001$) and \textit{Floral} ($p=$.020).
There was another main effect on emotional expression ($p$\textless.001). Pairwise comparisons revealed that \textit{Negative} ($m$=2.067 [1.661, 2.472]) had a lower affinity value than \textit{Positive} ($m$=4.704 [4.167, 5.241], $p$\textless.001) and \textit{Neutral} ($m$=4.254 [3.845, 4.663], $p$\textless.001). 

\begin{figure}[t]
  \centering 
  \includegraphics[width=.8\linewidth]{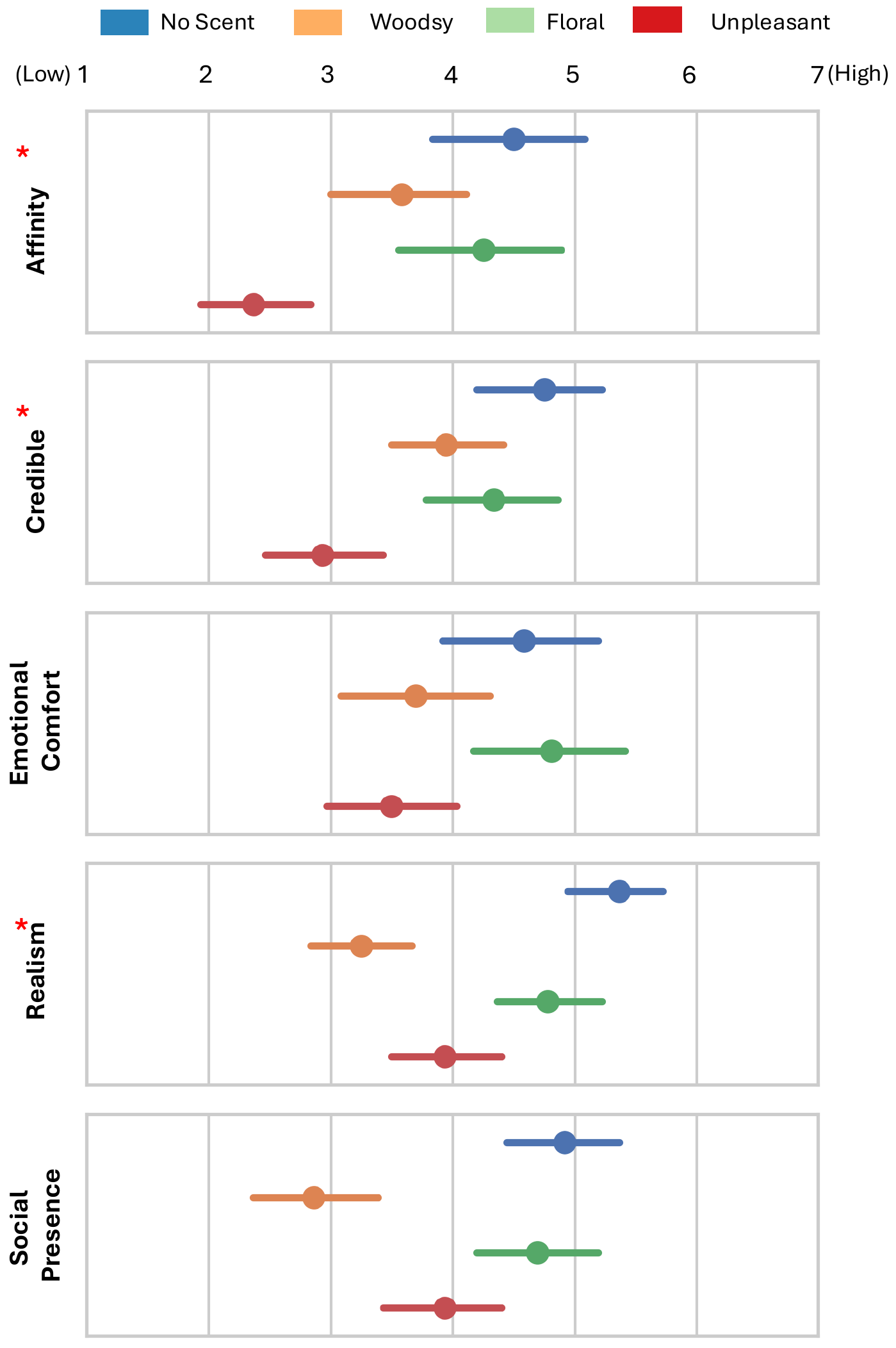}
  \vspace{-.3cm}
  \caption{
  Appeal - results by olfactory cues ranging from 1 to 7, with the error bars representing the 95\% confidence interval. We found the main effects of olfactory cues on Affinity, Credible, and Realism. 
  }
  \label{fig:appeal-credible-realism}
  \vspace{-.5cm}
\end{figure}

\paragraph{Credible:} We found a main effect on olfactory cues ($p$=.010). 
The olfactory cue results were \textit{No Scent}=4.75 [4.063, 5.437], \textit{Woodsy}=3.944 [3.258, 4.631], \textit{Floral}=4.333 [3.646, 5.020], and \textit{Unpleasant}=2.933 [2.181, 3.686] (Fig.~\ref{fig:appeal-credible-realism}). 
We found a significant difference between \textit{No Scent} and \textit{Unpleasant} ($p$=.008), indicating that \textit{Unpleasant} leads to lower perceived credibility compared to \textit{No Scent}.
We also found main effects on emotional expressions ($p$\textless.001) and on gender ($p$=.019).
Pairwise comparisons revealed that \textit{Negative} ($m$=2.983 [2.554, 3.413]) had significantly lower credibility than \textit{Positive} ($m$=4.571 [4.044, 5.098], $p$\textless.001) and \textit{Neutral} ($m$=4.417 [3.964, 4.870], $p$\textless.001). Next, \textit{Female} agent ($m$=4.217 [3.829, 4.604]) was perceived as more credible than \textit{Male} ($m$=3.764 [3.356, 4.171]). 

\paragraph{Emotional Comfort:} 
There was a significant main effect of emotional expression ($p$\textless.001). Pairwise comparisons showed that \textit{Negative} ($m$=3.208 [2.682, 3.735]) caused statistically more discomfort than \textit{Positive} ($m$=5.304 [4.479, 5.012], $p$\textless.001) and \textit{Neutral} ($m$=4.925 [4.479, 5.371], $p$=.011). 

However, no significant main effect of olfactory cue was found ($F(3, 19)$=2.312, $p$=.109, $\eta^2_p$=.267). The mean ratings for each olfactory condition were: No Scent=4.583 [3.722, 5.445], Woodsy=3.694 [2.833, 4.556], Floral=4.806 [3.944, 5.667], and Unpleasant=3.500 [2.556, 4.444].

\paragraph{Realism:} A significant interaction effect between emotional expression and gender was identified ($p$=.017). We found a simple effect on gender when participants interacted with the \textit{Positive} virtual agents ($p$=.013, $F(1, 22)$=7.366, $\eta^2_p$=.251). Participants perceived \textit{Positive Female} virtual agents ($m$=4.739 [4.153, 5.325]) as more real compared to \textit{Positive Male} ($m$=4.087 [3.398, 4.776]).


There was a main effect on olfactory cues ($p$\textless.001). 
Olfactory cue results were as follows: \textit{No Scent}=5.361 [4.389, 6.333], \textit{Woodsy}=3.250 [2.278, 4.222], \textit{Floral}=4.778 [3.806, 5.750], and \textit{Unpleasant}=3.933 [2.869, 4.998] (Fig.~\ref{fig:appeal-credible-realism}). Interestingly, only \textit{Woodsy} had significantly lower realism than \textit{No Scent} ($p$ =.027). 

A main effect was also observed for emotional expression ($p$=.032). The emotional expression results were as follows: \textit{Negative}=4.092 [3.516, 4.667], \textit{Positive}=4.408 [3.894, 4.923], and \textit{Neutral}=4.492 [3.990, 4.994]. Pairwise comparisons revealed a significant difference between \textit{Negative} and \textit{Neutral} ($p$=.008), but not between \textit{Negative} and \textit{Positive} ($p$=.269).

\paragraph{Social Presence:} We found a main effect on gender ($p$=.045). Participants perceived a statistically higher degree of social presence with \textit{Female} virtual agents ($m=$4.239 [3.542, 4.936]) than \textit{Male} ($m$=3.964 [3.329, 4.599]).

However, we found no significant main effect of olfactory cue ($F(3, 19)$=2.317, $p$=.108, $\eta^2_p$=.268). The results were No Scent=4.917 [3.642, 6.191], Woodsy=2.861 [1.586, 4.136], Floral=4.694 [3.420, 5.969], and Unpleasant=3.933 [2.537, 5.330].

\bgroup
\def\arraystretch{1.2}
\begin{table}[t]
\centering
\small
\caption{\label{table:big5_results} \revision{Big 5 Results: Significant interaction and main effects}}
\begin{tabular} {m{1.3cm} | m{2cm} | m{1.7cm} | m{0.6cm}| m{0.6cm}}
\toprule
\textbf{} & \textbf{Factor}  &  {$\bm{F}$} & {$\bm{p}$} & {$\bm{\eta_p}^2$}\\
\midrule
\textbf{Extraversion}   & {Emot. Expression}      &  {F(2, 38)=18.511}  &  {<.001} &  {.493}  \\ 
\midrule
                        & {Emot. Expression \newline \texttimes{} Gender}     &  {F(2, 38)=3.453} & {.045}  &  {.150}  \\ \cline{2-5}
  \textbf{Openness}   & {Emot. Expression}   &   {F(2, 38)=7.833}  & {.001}  & {.292}    \\ \cline{2-5}
                        &  {Gender}      &  {F(1, 19)=8.233}  & {.010}  & {.302}   \\
\midrule
                    & {Olfactory Cue}      &  {F(3, 19)=5.013}  &  {.010} &  {.442}  \\ \cline{2-5}
\textbf{Credible}   & {Emot. Expression}      &  {F(2, 38)=22.728} &  {<.001} &   {.545} \\ \cline{2-5}
                    & {Gender}      & {F(1, 19)=6.566}  & {.019}  &  {.257} \\ 
\midrule
\textbf{Conscienti- \newline ousness}   & {Emot. Expression}      &  {F(1.406, 26.719) =9.218}  &  {.002} &  {.327}  \\ 
\midrule
                       & {Emot. Expression \newline \texttimes{} Gender}      &  {F(2, 38)=8.374}  &  {.001} &  {.306}  \\ \cline{2-5}
\textbf{Agreeable- \newline ness}                        & {Olfactory Cue}      &  {F(3, 19)=3.286} &  {.043} &   {.342} \\ \cline{2-5}
                         & {Emot. Expression}      &  {F(2, 38)=50.545}  & {<.001}  &  {.727}  \\ 
\midrule
\textbf{ }              & {Olfactory Cue \newline \texttimes{} Emot. Expression}      &  {F(6, 38)=2.771}  &  {.025} &  {.304}  \\ \cline{2-5}
\textbf{Neuroticism}              & {Emot. Expression \newline \texttimes{} Gender}      &  {F(2, 38)=5.317}  &  {.009} &  {.219}  \\ \cline{2-5}
\textbf{}  & {Olfactory Cue}      &  {F(3, 19)=6.124} &  {.004} &   {.492} \\ \cline{2-5}
                         & {Emot. Expression}      &  {F(2, 38)=18.123}  & {<.001}  &  {.488}  \\ 

\bottomrule
\end{tabular}
\vspace{-2ex}
\end{table}

\subsection{Big Five Analysis Results}

\begin{figure}[t]
  \centering 
  \includegraphics[width=.8\columnwidth]{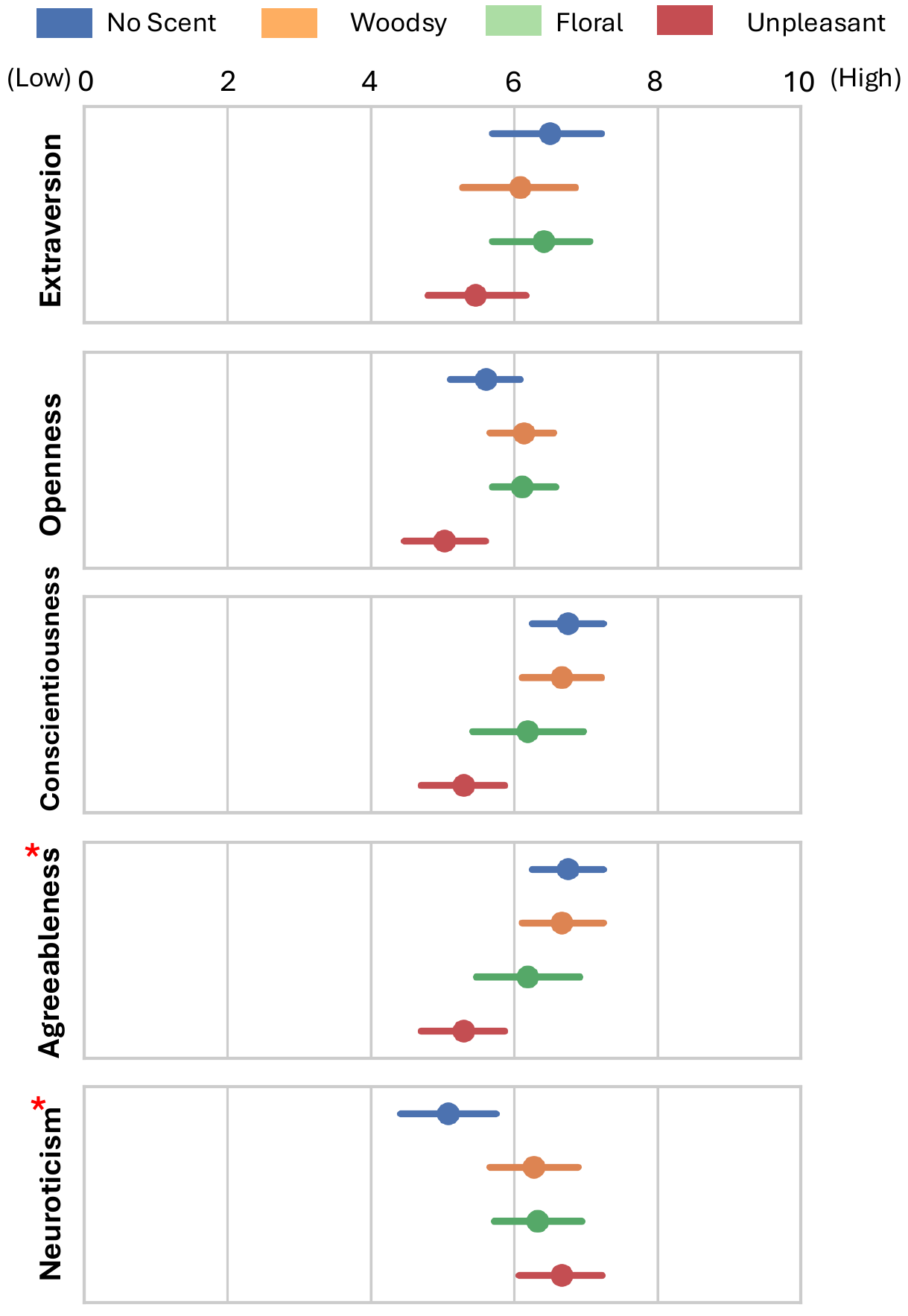}
  \vspace{-.3cm}
  \caption{   Big Five Traits - results by olfactory cues ranging from 0 to 10, with the error bars representing the 95\% confidence interval. The main effects of olfactory cue on Agreeableness and Neuroticism were disclosed. 
  }
  \label{fig:big5-agreeableness}
  \vspace{-.5cm}
\end{figure}

\revision{Table ~\ref{table:big5_results} shows the observed interaction and main effects in the Big 5 analysis results.}

\paragraph{Extraversion:} 
We discovered a main effect on emotional expression ($p$\textless.001). Pairwise comparisons showed that \textit{Positive} ($m$=7.533 [6.783, 8.284]) had a higher value than \textit{Neutral} ($m$=5.113 [4.451, 5.774], $p$\textless.001) and \textit{Negative} ($m$=5.704 [5.284, 6.124], $p$\textless.001). 
No significant difference between \textit{Neutral} and \textit{Negative} was found. 

No significant main effect of olfactory cue was disclosed ($p$=.224, $F(3, 19)$=1.594, $\eta^2_p$=.201) among No Scent ($m$=6.500 [5.767, 7.233]), Woodsy ($m$=6.083 [5.350, 6.817]), Floral ($m$=6.417 p[5.683, 7.150), and Unpleasant ($m$=5.467 [4.663, 6.270]).

\paragraph{Openness:} 
An interaction effect between emotional expression and gender was disclosed ($p=$.045). We found a simple effect on gender when the emotional expression is Positive ($p$<.001, $F(1, 22)$=18.104, $\eta^2_p$=.451). In Positive, \textit{Female} ($m$=6.739 [6.231, 7.248]) exhibited a higher degree of openness than \textit{Male} ($m$=5.478 [4.841, 6.115]).

We also found main effects on emotional expression ($p$=.001) and gender ($p$=.010). 
Pairwise comparisons of the emotional expression results revealed that \textit{Negative} ($m$=5.050 [4.455, 5.645]) had significantly lower openness than \textit{Positive} ($m$=6.071 [5.603, 6.538], $p$=.013) and \textit{Neutral} ($m$=6.05 [5.629, 6.471], $p$=.012). 
Next, \textit{Female} ($m$=6.025 [5.648, 6.402]) demonstrated greater openness than \textit{Male} ($m$=5.422 [4.975, 5.870]).

However, no significant main effect of olfactory cue was disclosed ($p$=.113, $F(3, 19)$=2.275, $\eta^2_p$=.264). The observed results were No Scent ($m$=5.611 [4.927, 6.295]), Woodsy ($m$=6.139 [5.455, 6.823]), Floral ($m$=6.111 [5.427, 6.795]), and Unpleasant ($m$=5.033 [4.284, 5.783]).

\paragraph{Conscientiousness:} We found a main effect on emotional expression ($p$=.002). \textit{Negative} ($m$=5.329 [4.707, 5.952]) had significantly lower conscientiousness than \textit{Positive} ($m$=6.854 [6.167, 7.542], $p$=.014) and \textit{Neutral} ($m$=6.500 [6.045, 6.955], $p$=.006).

No significant main effect of olfactory cue was disclosed ($p$=.060, $F(3, 19)$=2.933, $\eta^2_p$=.317). The observed results were No Scent=6.750 [5.984, 7.516], Woodsy=6.667 [5.901, 7.432], Floral=6.194 [5.429, 6.960], and Unpleasant=5.300 [4.461, 6.139].

\paragraph{Agreeableness:} 
An interaction effect between emotional expression and gender was reported ($p$=.001).
A simple effect on gender was observed with Positive ($p$=.005, $F(1, 22)$=9.708, $\eta^2_p$=.306), indicating \textit{Female} ($m$=7.435 [6.665, 8.204]) had a higher score than \textit{Male} ($m$=6.304 [5.445, 7.164]).

A main effect on olfactory cue ($p$=.043) was found. The observed results were \textit{No Scent}=6.361 [5.583, 7.140], \textit{Woodsy}=5.472 [4.694, 6.251], \textit{Floral}=5.556 [4.777, 6.334], and \textit{Unpleasant}=4.633 [3.781, 5.486]. It is shown in Fig~\ref{fig:big5-agreeableness}. We only discovered a significant difference between \textit{No Scent} and \textit{Unpleasant} ($p$=.033).
There was a main effect on emotional expression ($p$\textless.001)
Pairwise comparisons revealed that \textit{Negative} ($m$=3.613 [3.146, 4.079]) had significantly lower agreeableness than \textit{Positive} ($m$=6.804 [6.150, 7.458], $p$\textless.001) and \textit{Neutral} ($m$=6.100 [5.535, 6.665], $p$\textless.001). 


\begin{figure}[t]
  \centering 
  \includegraphics[width=.8\linewidth]{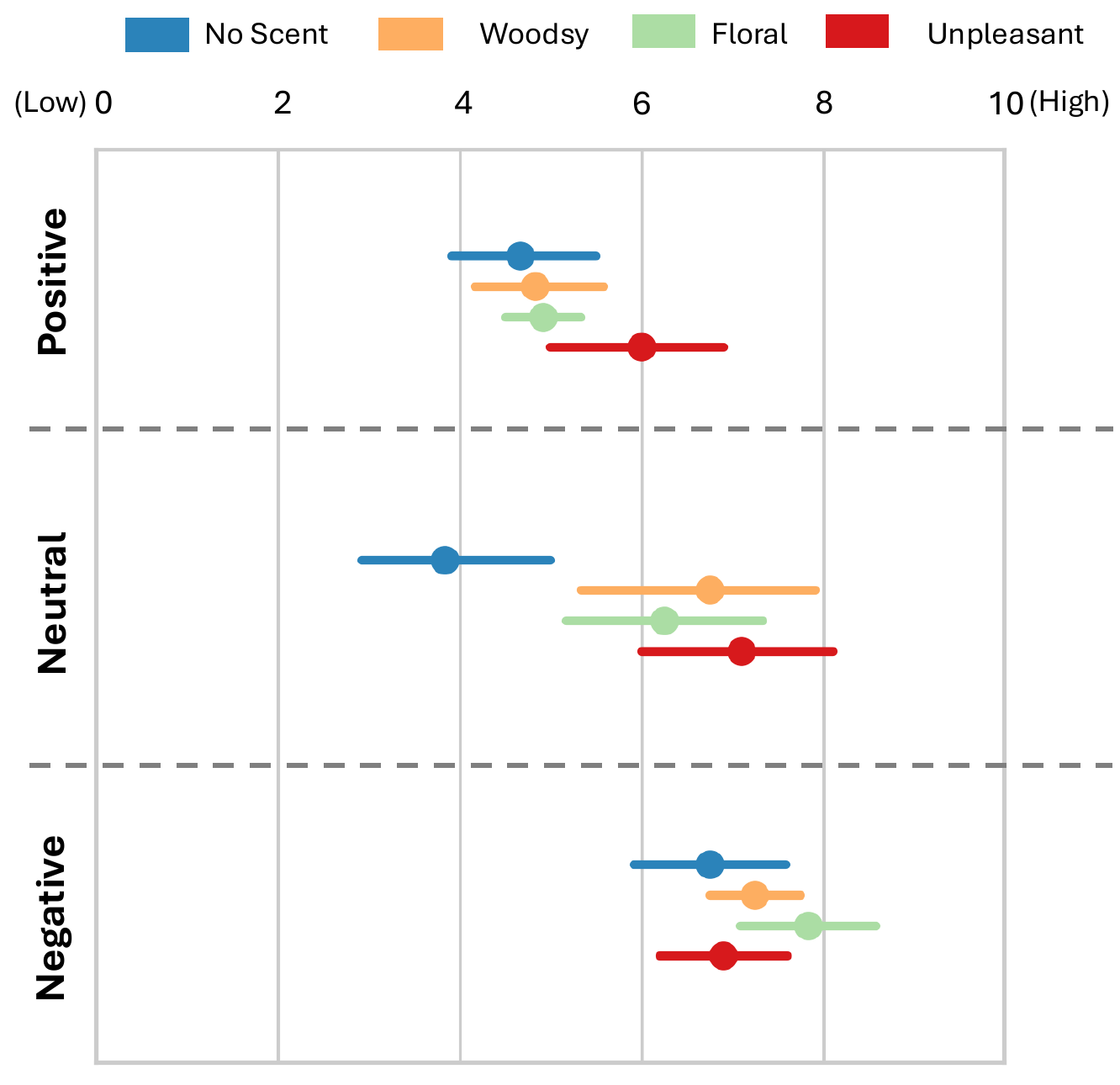}
  \vspace{-.3cm}
  \caption{Big Five Traits - Neuroticism results ranging from 0 to 10. The error bars represents the 95\% confidence interval. An interaction effect between emotional expression and olfactory cue was found. 
  }
  \vspace{-.3cm}
  \label{fig:big5-neuroricism}
\end{figure}

\paragraph{Neuroticism:} Two interaction effects between emotional expressions and olfactory cue ($p$=.025) and between emotional expressions and gender ($p$=.009) were found. A simple effect on the olfactory cue is found when the virtual agent is \textit{Neutral} ($p$=.002, $F(3, 45)$=5.921, $\eta^2_p$=.297) as shown in Fig~\ref{fig:big5-neuroricism}. \textit{No Scent} ($m$=3.833 [2.597, 5.070]) has significantly lower neuroticism compared to \textit{Woodsy} ($m$=6.750 [5.263, 8.237], $p$=.008), \textit{Floral} ($m$=6.250 [4.919, 7.581], $p$=.039), and \textit{Unpleasant} ($m$=7.100 [5.818, 8.382], $p$=.004). 
There was a simple effect on gender for the \textit{Neutral} virtual agent ($p$=.011, $F(1, 22)$=7.622, $\eta^2_p$=.257). \textit{Female} ($m$=6.696 [5.665, 7.726]) had a higher score than \textit{Male} ($m$=5.174 [4.237, 6.111]).

There was a main effect on olfactory cues ($p$=.004). 
Pairwise comparisons showed that \textit{No Scent} ($m$=5.083 [4.509, 5.657]) had statistically lower score than \textit{Woodsy} ($m$=6.278 [5.704, 6.852], $p$=.037), \textit{Floral} ($m$=6.333 [5.759, 6.907], $p$=.027), and \textit{Unpleasant} ($m$=6.667 [6.038, 7.296], $p$=.006). 
The main effect on emotional expression ($p$\textless.001) was revealed. 
Pairwise comparisons showed that \textit{Negative} ($m$=7.183 [6.742, 7.624]) had significantly higher score than \textit{Neutral} ($m$=5.983 [5.366, 6.601], $p$=.008) and \textit{Positive} ($m$=5.104 [4.645, 5.563], $p$\textless.001). However, there was no significant difference between \textit{Neutral} and \textit{Positive}.






\subsection{Biosignal Analysis Results}

We analyzed biosignals collected from 21 out of 23 participants, excluding two (one in the \textit{No Scent} group and another in the \textit{Woodsy} group) who had trouble in recording their biosignal data. ANOVA results revealed a main effect on the virtual agent's gender ($p$=.04), while no significant effects were found for the agent's emotional expression or olfactory cues.  
It quantitatively showed that participants had a more positive impression of the \textit{Female} virtual agent ($\Delta$$pNN50=8.906ppt$) than \textit{Male} ($\Delta$$pNN50$=4.695ppt).

\subsection{Subjective Results}

After participants completed the entire study, 
they were asked to evaluate the effect of emotional expression and olfactory cue on their judgment of female and male virtual agents, respectively, using a 7-point Likert scale (1: low - 7: high). 
The results are shown in Fig.~\ref{fig:subjective-rate}. 
For both female and male virtual agents, most participants reported that emotional cues from facial expressions and behavior strongly influenced their judgments, with the majority giving high ratings (5–7). 
In contrast, the effect of olfactory cue on their judgment was mixed and less consistent. While some participants perceived scent as highly influential, others rated it much lower, suggesting that the impact of olfactory cue was more subjective and less reliable compared to visually observable emotional expressions. We discuss this further in Sec.~\ref{sec_discussion}.

\begin{figure}[t]
  \centering 
  \includegraphics[width=.8\linewidth]{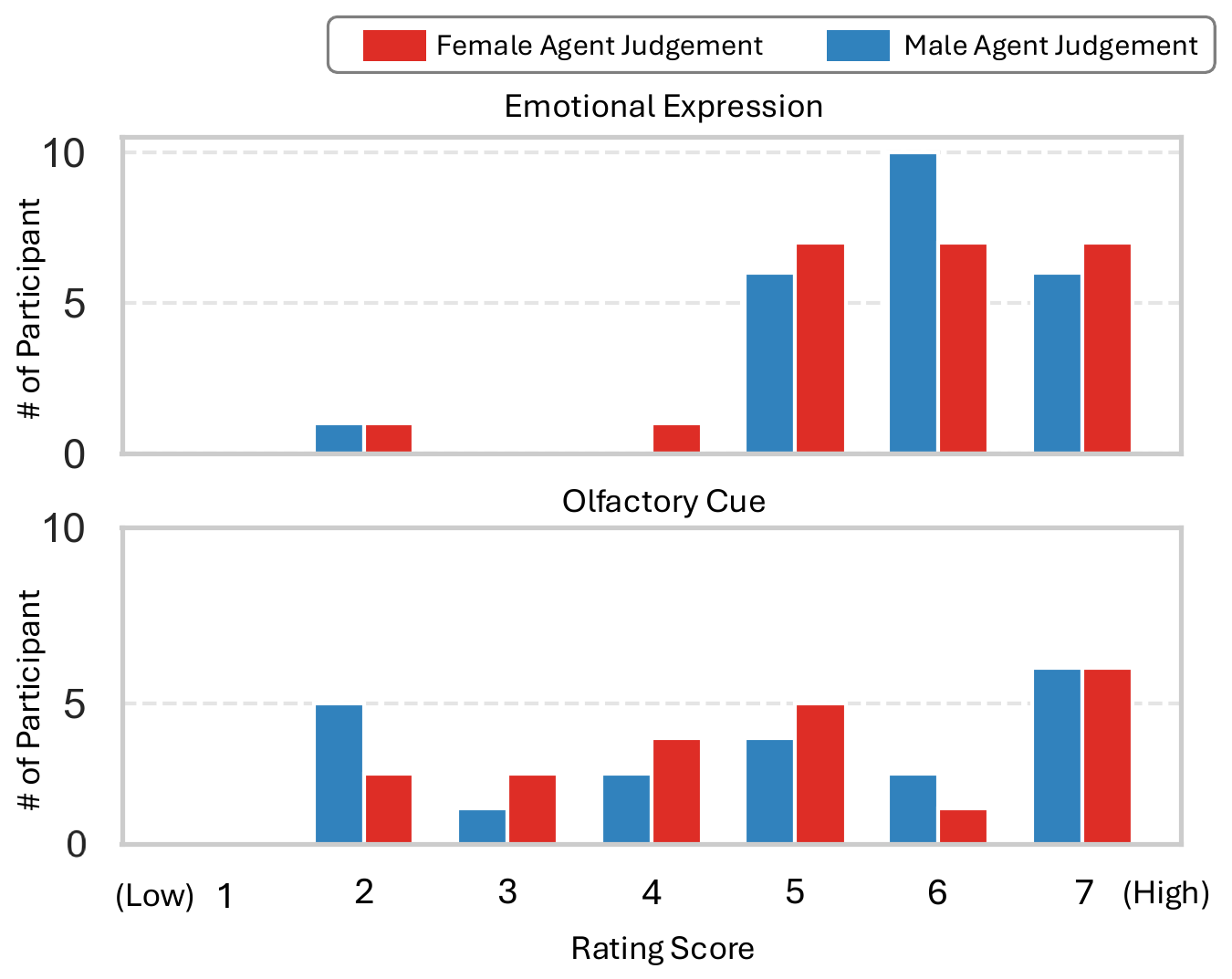}
  \vspace{-.3cm}
  \caption{
  \revision{Participants’ subjective ratings of the influence of emotional expressions and olfactory cues on their judgments of female and male virtual agents. Participants largely relied on emotional cues, with most assigning high ratings (5–7), whereas responses to olfactory cues were more varied and less consistent.
  }
  }
  \label{fig:subjective-rate}
  \vspace{-.5cm}
\end{figure}

\subsection*{Summary of Analysis Results}
First, we found no evidence supporting \textbf{H1}, as there were no clear differences among \textit{Woodsy}, \textit{Floral}, and \textit{No Scent}. 
An exception was observed for neuroticism, where the \textit{No Scent} condition yielded significantly lower scores compared to both scented conditions. 

Second, our results support \textbf{H2}, showing that \textit{Unpleasant} has a significantly negative effect on users’ judgments of virtual agents’ impressions.
Third, we rejected \textbf{H3}, as no clear evidence of improvement was found when using scented conditions compared to the \textit{No Scent} condition. Contrary to our expectation, participants reported lower realism in the \textit{Woodsy} condition than in the \textit{No Scent} condition. Additionally, we found no interaction effect between olfactory cues and agent gender, leading us to reject \textbf{H4}. 

Next, our results partially support \textbf{H5} as \textit{Positive} and \textit{Neutral} were perceived more positively than \textit{Negative}. However, no clear difference between \textit{Positive} and \textit{Neutral} was found.
\textbf{H6} is also partially supported. 
Our results demonstrate that when virtual agents' emotional expressions were identical, the \textit{Female} virtual agent was perceived as more positive and appealing than the \textit{Male} virtual agent. This finding is also supported by the biosignal results.

\section{Discussion}\label{sec_discussion}

\subsection{Predicting User Engagement with Virtual Humans}




To predict participants' future engagement with virtual agents, a multivariate logistic regression model was employed. We used Python package scikit-learn (version 1.2.1). This is aim to examine how the Big Five and appeal metrics influence participants' decisions to engage in future interactions with virtual agents. Accordingly, the model incorporated the Big Five personality traits and appeal metrics as predictors,with participants’ responses to the future engagement question serving as the outcome variable. 
The confidence level for the regression model was set at 99\%. 

Among the predictors, affinity ($\beta$=1.75, $p<.001$) was found to be a significant positive predictor, while other predictors did not reach statistical significance. This suggests that higher affinity significantly increases the likelihood of future engagement, while other factors did not show a meaningful effect in this model. The affinity values range from 1 to 7.
The model's predictive equation is as follows, where -9.61 represents the intercept ($p=.005$):
\[
P(\text{futureEngagement} = 1) = \frac{1}{1 + e^{-(-9.61 + 1.75 \times \textit{affinity})}}
\]

\subsection{Summary of Findings}
We discuss our findings based on the results and participant feedback, followed by a discussion of study limitations and future work.

\paragraph{\textbf{Olfactory cues influenced how participants perceive virtual agents, reflecting real-world patterns of social perception influenced by olfactory cues.}}
Our results showed a clear preference for the \textit{No Scent} condition over \textit{Unpleasant}, but no statistical evidence that \textit{Woodsy} and \textit{Floral} were superior to \textit{No Scent}. 
In our regression analysis, we identified affinity as the strongest positive predictor of participants' willingness to engage in future interactions with virtual agents. 
Additionally, as discussed in Section~\ref{Sec:AppealAnalysisResults}, our results demonstrated a main effect of olfactory cues, indicating that both \textit{No Scent} and \textit{Floral} contributed to higher levels of affinity compared to \textit{Unpleasant}. This suggests that while pleasant scents may have a minor enhancing effect, avoiding unpleasant scents is crucial for improving the perceived affinity of virtual agents. 
Participants' feedback supports this as well.


Our results are consistent with earlier psychology research in real-world settings, which found 1) individuals rated faces as less attractive in the presence of unpleasant scents compared to clean air and pleasant scents~\cite{dematte2007olfactory};
2) unpleasant scents have a greater impact on attractiveness ratings compared to pleasant scents when assessed against a baseline (pure air)~\cite{cook2018simultaneous}; and
3) no clear difference in estimating the subjective emotions of faces when presented with pleasant scents compared to no scents~\cite{bensafi2002modulation}. These consistent findings suggest that participants' judgments of virtual agents may, at least in part, mirror the way we evaluate others in the real world. To gain deeper insights, further research involving a variety of scents and accounting for individual scent preferences is necessary.

\paragraph{\textbf{The emotional expressions of virtual agents had the greatest influence on their impressions.}} 
This finding suggests that impressions of the virtual agents are formed more quickly using visually observable cues rather than olfactory cues. 
Participants remarked that their initial impression of a virtual agent was primarily influenced by the agent's emotional expression. One participant reported feeling more relaxed when the agents smiled. 
She added that once she formed a positive impression based on facial expressions, the olfactory cue played only a minor role in reinforcing her perception. However, if her initial impression was negative, she completely disregarded the olfactory cue. 
Regarding the Negative virtual agents, the participants reported feeling attacked and perceiving the agents as less friendly. 
In contrast, they found positive expressions and behaviors made the agents appear more trustworthy and sociable.
These findings are consistent with previous research~\cite{bonsch2020impact}, which found that participants adjust their interpersonal distance from a virtual agent based on facial expressions.


\paragraph{\textbf{The female virtual agent was perceived as more positive or attractive than the male agent.}} 
This finding is supported not only by our results but also by participant comments.
One participant commented that even when both virtual agents appeared to be friendly (Positive), the female character felt far more trustworthy. 
This comment is aligned with the earlier study~\cite{guadagno2007virtual, zibrek2020effect}, confirming the women-are-wonderful effect. 
Moreover, this finding is also aligned to previous interpersonal distance studies~\cite{bailenson2003interpersonal, brady1978interpersonal, polit1977sex}, having smaller interpersonal distances to female agents than male agents. 
However, it is worth noting that our study had a higher proportion of female participants (14) than male participants (9). 
This population imbalance in this study makes it challenging to investigate the gender effect on impressions of virtual humans, known as in-group bias.



\subsection{Considerations: Olfactory Cues and Virtual Agents}

Participants' feedback supported the nature of olfactory perception in VR is complex and subjective, just as it is in the real world. This might partially explain reasons behind the rejection of our \textbf{H1}.

Individual differences in scent sensitivity and personal preferences may lead to varied responses to virtual agents, influencing participants' perceptions and interactions. Two participants provided particularly insightful perspectives on this issue.
While pleasant scents (\textit{Floral} and \textit{Woodsy}) are often associated with good hygiene and extroversion, they can also be perceived as intrusive when experienced in close proximity, creating discomfort or psychological distance between users and virtual agents.
One participant in the \textit{Woodsy} group mentioned that she was overwhelmed by the intensity of the scent.
She associated it with the virtual agent's strong personality, interpreting it as domineering or overbearing, and expressed a desire to distance herself from the agent. \revision{This highlights the need to carefully regulate scent intensity when incorporating olfactory cues in virtual agent design.}
Similarly, another participant reported a dislike for \textit{Floral}, stating that it made him uncomfortable and that he did not want to remain in the room with the virtual agent for an extended period.
This finding reinforces the need for personalized or adaptive olfactory cues in VR to optimize user comfort and interaction quality with virtual agents.


We also identified another interesting consideration for the effective use of olfactory cues with virtual agents.
Participants' real-world experiences appeared to influence how they perceived virtual agents, suggesting that scent-based cues can evoke personal memories and emotional responses that shape user interactions. This supports earlier findings~\cite{chanes2016redefining, sullivan2015olfactory}.
Two participants specifically highlighted this effect. 
One participant, who was exposed to the \textit{Floral} scent, reported that the male virtual agent reminded her of someone she knew in real life. 
Her previous association shaped how she perceived the virtual agents, showing that scent can be a powerful cue for memory recall and can significantly influence users’ impressions of virtual agents.
Another participant, who experienced the \textit{Unpleasant} scent, noted that it elicited negative memories or emotions, which in turn transferred to the virtual agents. This response suggests that aversive olfactory cues can unintentionally create negative biases toward virtual agents, potentially reducing user engagement and comfort in VR interactions. This finding highlights the deep connection between scent, memory, and emotion. 

\subsection{Limitation and Future Work}
One limitation is that the impact of olfactory cues on long-term interactions with virtual agents remains unclear. 
While our findings suggest that olfactory cues—particularly the \textit{Unpleasant} scent—significantly influenced users' judgment of virtual agents, it remains unclear whether these effects persist during prolonged interactions with virtual agents. 
Additionally, it remains unexplored whether olfactory cues have similar effects on user judgment when users encounter a virtual avatar instead of a virtual agent.

\revision{Another limitation lies in the use of the same male and female virtual agent models (i.e., identical face and attire) across all conditions. While this choice was made to control for visual consistency, it may have introduced an unintended order effect, where early-formed impressions persisted and influenced responses in later trials. 
}
Future research should investigate a broader range of scent types and their compatibility with various characteristics of virtual humans, including age, race, gender, and attire. 

\revision{Lastly, the participant sample in this study was limited in both size and diversity. Future research would benefit from engaging a larger and more diverse population to strengthen and broaden the generalizability of the findings.}
Moreover, users from diverse backgrounds—varying in age, race, and culture—may perceive scents differently. Scents may evoke different associations or biases. Understanding these cross-cultural and individual differences will be critical for enhancing user satisfaction and inclusivity, ensuring that olfactory technologies are adaptable for a wide range of users.
These future works should be investigated in conjunction with the development of advanced olfactory devices capable of delivering a wider variety of scents and controlling their intensity.

\section{Conclusion}

This study explored how olfactory cues, emotional expressions, and the gender of virtual agents influence users’ judgments and impressions of those agents. We found that emotional expression had the most influence, suggesting that virtual agents should avoid negative emotional expressions. 
Furthermore, when the virtual agents had the same emotional expressions, the female agents were more positively perceived than the male agents.
While we found no significant difference among \textit{No Scent}, \textit{Woodsy}, and \textit{Floral} olfactory cues, it is recommended to avoid unpleasant scents.
\revision{More specifically, our findings indicate that participants formed negative impressions of virtual agents that emitted unpleasant scents, regardless of emotional expressions or gender. Conversely, under \textit{No Scent}, \textit{Woodsy}, and \textit{Floral} conditions, participants relied more heavily on visual cues to form their judgments about virtual agents.
These findings appear consistent with prior research on human judgment via olfactory cues, tentatively suggesting participants may judge virtual agents similarly to real-world individuals.}
Furthermore, we gained insights from participant feedback that olfaction can have varying effects depending on individual experiences and preferences.

\newpage

\bibliographystyle{abbrv}

\bibliography{0_main}
\end{document}